\input phyzzx.tex

\def\qgg {\Delta_{\hat g}}

\def\d {\partial}
\def\v {\varphi}
\def\l {\lambda}
\def\a {\alpha}
\def\g {\gamma}
\def\z {\bar z}
\def\t {\theta}
\def\f {\phi}
\def\b {\beta}
\def\G {\Gamma}
\def\F {\Phi}
\def\picc {\scriptscriptstyle}
\def\s {\sigma}
\def\e {\epsilon}
\def\D {\Delta}
\def\dd {\delta}
\def\L {\Lambda}
\def\ze {\zeta}

\titlepage
\title{The Quantization of Anomalous Gauge Field Theory and BRST-invariant
Models of Two Dimensional Quantum Gravity}

\author {M. Martellini\foot{On leave of absence from Dipartimento di Fisica, 
Universit\'a
di Milano, Milano, Italy and I.N.F.N., Sezione di Milano, Italy}}
\address{The Niels Bohr Institute, University of Copenhagen, Copenhagen $\phi$,
 Denmark and Landau Network at Centro Volta, Como, Italy}
\author {M. Spreafico}
\address{Dipartimento di Matematica, Universit\`a di Milano, Milano
and I.N.F.N., Sezione di Milano, Italy}
\author {K. Yoshida}
\address{Dipartimento di Fisica, Universit\`a di Roma, Roma and
I. N. F. N., Sezione di Roma, Italy}
\abstract{We analyze the problems with the so called gauge invariant 
quantization of the anomalous gauge field theories originary due to
Faddeev and Shatashvili (FS). Our analysis bring to a generalization
of FS method which allows to construct a series of classically equivalent
theories which are non equivalent at quantum level.  
We prove that these classical theories are all consistent with the
BRST invariance of the original gauge symmetry with suitably augmented
field content. As an example of such a scenario, we discuss the class of
physically distinct models of two dimensional induced gravity which are a 
generalization of the David-Distler-Kawai model.}

\endpage

\pagenumber=1

\vglue 0.6cm
{\bf 1. Introduction}
\vglue 0.4cm
The consistent quantization of (classical) gauge invariant field theory 
requires the complete cancellation of anomalies [1] [2]. Here, "consistent"
means that we want not only to require
renormalizability (perturbative finiteness), but also 
unitarity of S-matrix, non-violation of Lorentz invariance etc.
Moreover, in physical $4d$ world, anomaly cancellation condition itself 
often leads to the physical predictions. The well known example is the equality
of numbers of quarks and leptons in the Standard Model of Weinberg and Salam.

On lower dimensional (eg. $d=2$) field theory, the cancellation of 
anomalies is still the crucial ingredient for the model building. The critical 
string dimension $d=26$ is often quoted [3] as the consequence of anomaly 
free condition for bosonic string (although in this example the cancellation
of anomaly does not guarantee full consistency of the model 
in above sense, due to
the presence of tachyons).

In the case of lower dimensional field theory ($d<4$), one often 
tries to
quantize a gauge field theory when there is no way of cancelling
its anomaly. The classical example of this situation 
is the attempt to the quantization of chiral
Schwinger model by Jackiw and Rajaraman [4] [5]. They have shown that the 
model can be consistently quantized (=free field theory) even when the gauge
invariance is broken through anomaly.

In general there seem 
to be two ways for attempting the quantization of anomalous 
gauge field theory:

\noindent{1) {\it Gauge non invariant method}}

One ignores the breaking of 
gauge symmetry and try to show that the theory can be 
quantized even without the gauge invariance. The example of this approach
is the above Jackiw-Rajaraman quantization of the chiral Schwinger model. The 
problem here is that it is not easy to develop the general technics covering
wide class of physically relevant models with anomaly.

\noindent{2) {\it Gauge invariant method}}

In this case, one first tries to recover gauge invariance
by introducing new degrees of freedom. The theory is anomalous
when one can not find local counter term to cancel the gauge non invariance
due to the one loop "matter" integrals in presence of gauge fields,
by making use
exclusively of the degrees of freedom (fields) already present in the classical
action.

In ref. [6], Faddeev and Shatashvili (FS) have tried to justify the 
introduction of new degrees of freedom which are necessary to construct the
anomaly cancelling counter term. Their argument is based on the idea of
projective representation of gauge group. 
They observe that the appearance of anomaly does 
not mean the simple breakdown of (classical) gauge symmetry, 
but it rather signals
that the symmetry is realised projectively (this is related to the appearance
of anomalous commutators of relevant currents). Such a realization,
through projective representations, necessitates the enlargement of 
physical Hilbert space. Thus they argued that the introduction
of new fields in the model is not an ad hoc (and largely arbitrary) 
construction.

Independently of their "philosophy", the FS method gives the gauge 
invariant action at the price of introducing the extra degrees of freedom
(generally physical). The serious problem of this method is, however, that 
the gauge invariance thus "forced" upon the theory, does not automatically
guarantee the consistency of the theory. This is in contrast with our
experience with some $4d$ models such as the Standard Model.

For example, one may apply the FS method to the celebrated case of chiral
Schwinger model [4] [5] [5A]. In this case, we have the classical action
$$
S_{\picc 0}= \int {dz\wedge d\z\over 2i}
\Big [\bar\psi_{\picc R}\g_{\z}(\bar\d+R)\psi_{\picc R}+
\bar\psi_{\picc L}\g_z\d\psi_{\picc L}+{1\over 4} Tr~F^2\Big]
$$
where
$$
\eqalign{&\psi_{\picc R/L}={1\pm \g_5\over 2}\psi\cr
&R/L=A_1\pm iA_2, ~F=\bar\d L-\d R+[R,L]\cr}
$$
(we are using the euclidian notation).

So is invariant under the gauge transformation 
$$
\eqalign{
&\psi_{\picc R}\to \psi_{\picc R}^g=S(g) \psi_{\picc R}\cr
&\psi_{\picc L}\to\psi_{\picc L}\cr
&A_\mu=g A_\mu g^{-1}+g\d_\mu g^{-1}\cr}
$$
for any $g(z,\z)\in G$.

The theory is anomalous because the one loop integral
$$
e^{-W_{\picc R}(R)}=\int {\cal D}\psi_{\picc R}{\cal D}\bar\psi_{\picc R}
exp-\int\bar\psi_{\picc R}\g_{\z} (\bar\d+R)\psi_{\picc R}
$$
is not gauge invariant under
$$
R\to g R g^{-1}+g\bar\d g^{-1}
$$
(for any choice of the regularization).

Following FS'technic (see next section) however, one can introduce the local
counter term $\L(R, L;g)$, $(g(z,\z)\in G)$ so that the gauge variation
of $\L$ cancels the non invariance of $W_{\picc R}(R)$.

There is certain arbitrariness in the choice of $\L$ but the convenient one
is 
$$
\L(R,L;g)=-\Big ( \a_{\picc L}(L,g) + {1\over 4\pi} \int Tr(RL)\Big)
$$
where
$$
\eqalign{
&\a_{\picc L}(L,g)={1\over 4\pi}\Big[ -\int {dz\wedge d\z\over 2i} Tr(g^{-1}
\bar\d g,L)+{1\over 2}\int {dz\wedge d\z\over 2i} 
Tr(g\d g^{-1},g\bar\d g^{-1})\cr
&-{1\over 2}\int_0^1dt\int {dz\wedge d\z\over 2i}Tr(g'\d_tg'^{-1},[g'\d g'^{-1},
g'\bar\d g'^{-1}])\Big]\cr
&g'(0,z,\z)=1,~~~g'(1,z,\z)=g(z,\z)\cr}
$$
is the Wess-Zumino-Novikov-Witten action corresponding to the anomaly of left
fermion $\psi_{\picc L}$, $\bar\psi_{\picc L}$ ($\a_{\picc L}$ is not globally
a local action but it is so for "small" $g\simeq 1+i\xi$). 
That is, one can write
$$
\a_{\picc L}(L,g)=W_{\picc L}(L^g)-W_{\picc L}(L)
$$
where
$$
e^{-W_{\picc L}(L)}=\int {\cal D} \psi_{\picc L}'{\cal D}\bar\psi_{\picc L}'
exp-\int \bar\psi_{\picc L}'\g_z (\d+L)\psi_{\picc L}'
$$
(note that $\psi_{\picc L}',\bar\psi_{\picc L}'$ have nothing to do with 
$\psi_{\picc L},\bar\psi_{\picc L}$ in $S_{\picc 0}$).

With this choice of counter term, one can show that the theory is equivalent
to {\it a}) free decoupled fermion $\psi_{\picc L},\bar\psi_{\picc L}$ and 
{\it b}) the vector Schwinger model. In fact, the added bosonic degree of
freedom $g(z,\z)\in G$ can be "fermionized" to act as missing $\psi_{\picc L}',
\bar\psi_{\picc L}'$ with the 
right coupling to the left component $L$ of gauge field.

However, there is still a point missing in this story. In fact, after 
introducing the new degree of freedom $g$, there is no reason to exclude
the other type of invariant local counter term such as
$$
{a\over 4\pi}\int Tr (L^g R^g)=
{a\over 4\pi}\int Tr\Big( (g L g^{-1} + g \d g^{-1}), (g R g^{-1}
+g \bar\d g^{-1})\Big)
$$
(one can also attribute it to indefinite - regularization dependent -
part of fermionic integral, i.e. $W_{\picc R}(R)+W_{\picc L}(L) + {a\over
4 \pi}\int Tr (RL)$).

It is well known [4] that the arbitrary constant $a$ 
enters the physical spectrum.
For abelian case, $G=U(1)$, the mass square of massive boson is given by
$$
m^2={e^2 a^2\over a-1}
$$
thus, for $a<1$, the theory is not consistent although the requirement of gauge
invariance is satisfied.

In the fermionized version of the theory [5A], 
$a$ enters the charges of left and right fermions as
$$
e_{\picc R/L}={e\over 2}\Big( \sqrt{a-1}\pm {1\over \sqrt{a-1}}\Big)
$$

This means that the condition $a>1$ is necessary also for the real coupling 
constant, or the hermitian hamiltonian.

In general, the consistence of the theory can be proved if one can set
up the BRST scheme with certain physical conditions at the start, such as
hermiticity of the hamiltonian [11].

In what follows, we discuss the possibility of recasting the FS method into
BRST formalism, thus facilitating the analysis of the consistency
of the theory.

\vglue 0.6cm
{\bf 2. Faddeev-Shatashvili method}
\vglue 0.4cm
\noindent{\it a) Path integral formalism}

We shall briefly describe the Faddeev-Shatshvili (FS) method of quantizing
anomalous gauge field theory in the path integral formalism, following
the work of Harada and Tsutsui [7], Babelon, Shaposnik and Vialet [8].

Let us take a generic gauge field theory described by the classical action
$$
S_{\picc 0}(A,X)=S_{\picc G}(A)+S_{\picc M}(X;A)
\eqn\uno
$$
where $\{A(x)\}$ and 
$\{X(x)\}$ represent respectively gauge fields and "matter fields",
gauge invariantly coupled to the former.

The total action $S_{\picc 0}$ 
as well as the pure gauge part $S_{\picc G}$ and the matter
part $S_{\picc M}$ are invariant under the local gauge 
transformation
$$
\eqalign{&A\to A'=A^g\cr &X\to X'=X^g\cr}~~~~ g(x)\in G. 
\eqn\due
$$

Being anomalous 
generally means that the one loop matter integral (assumed that $S_{\picc M}
(X,A)$ is quadratic in  $X$)
$$
\int {\cal D}X e^{-S_{\picc M}(X,A)}\equiv e^{-W(A)}
\eqn\tre
$$
cannot be regularized in such a way as to preserve the gauge invariance
of the functional $W(A)$, 
$$
W(A^g)-W(A)=\a(A;g)\not= 0
\eqn\quattro
$$

Naturally, $\a(A;g)$ depends on the regularization used, 
but there is no way of cancelling it completely by adding 
some local counter
term $\L(A,X)$ to the action.

One can understand eq. \quattro\ as the non invariance of the path 
integral measure, ${\cal D} X$:
$$
{\cal D} X^g\not= {\cal D} X
\eqn\cinque
$$

In fact, as shown by Fujikawa [9], one can write the "anomaly equation"
$$
\eqalign{&W(A^g)-W(A)=\a(A;g)\cr
&det\left({{\cal D}X^g\over {\cal D} X}\right) = e^{-\a(A;g)}=
e^{\a(A;g^{-1})}\cr}
\eqn\sei
$$

In this situation, clearly one can not hope that the usual Faddeev-Popov
(FP) ansatz to quantize the theory may go through. 

If one inserts the $\delta$ function identity
$$
1=\D(A)\int {\cal D} g \delta(F(A^g))
\eqn\sette
$$
where $F(A)$ is a gauge fixing function, into the path integral
expression for the partition function 
$$
Z=\int {\cal D} A\int {\cal D}X e^{-(S_{\picc G}(A)+S_{\picc M}(X;A))}
$$
then one obtains
$$
\eqalign{Z=&\int {\cal D} A \int {\cal D} X \D (A) e^{-[S_{\picc G}(A)+
S_{\picc M}(X;A)]}
\int {\cal D} g \delta (F(A^g))\cr
=&\int {\cal D} A\int {\cal D} g\D (A) e^{-S_{\picc G}(A^g)}
\int {\cal D} X e^{-S_{\picc M}
(X^g;A^g)} \delta (F(A^g))\cr}
\eqn\otto
$$
the second equality follows from 
the gauge invariance of the classical action: $S_{\picc 0}(A^g;X^g)=
S_{\picc 0}(A;X)$.

In the case of usual gauge field theory, such as chiral Schwinger model, 
we can make a series of 
assumptions on the remaining functional measures ${\cal D} A$ and
${\cal D}g$.

First, we assume
$$
(1)~~~~~~~{\cal D} A={\cal D}A^g
\eqn\nove
$$
then with the change of variable $A^g\to A$ and $X^g\to X$ in \sette, we get
$$
Z=\int {\cal D} g  \int {\cal D} A\D \delta(F(A))(A^{g^{-1}})
e^{-S_{\picc G}(A)} 
\int {\cal D} X e^{-[S_{\picc M}(X;A)+\a(A;g^{-1})]}
\eqn\dieci
$$
where we have used eq. \cinque, i.e. ${\cal D}X={\cal D}X^{gg^{-1}}=
{\cal D}X^g e^{-\a(A;g^{-1})}$.

Further, one can assume for the usual gauge group, 
the invariance of Haar measure ${\cal D} g$, i.e. for any $h$ in $G$
$$
(2)~~~~~~~{\cal D}(gh)={\cal D}(hg)={\cal D} g
\eqn\undici
$$
which results, as is well known, in the invariance of the FP factor
$\D(A)$
$$
\D(A^{g^{-1}})=\D(A)
\eqn\dodici
$$

Thus, we get the expression for $Z$ proposed in ref.s [6] and [7]
$$
Z=\int {\cal D} g \int {\cal D} A\D (A)\delta(F(A))\int {\cal D} X
e^{-S_{\picc eff}(X,A;g)}
\eqn\tredici
$$
with
$$
S_{\picc eff}(X,A;g)=S_{\picc 0}(X,A;g)+\a(A;g^{-1})
\eqn\quattordici
$$

As one can see from eq. \quattro\ the effect of the counter term 
$\a(A;g^{-1})$ is to transform the one loop path integral $W(A)$, eq. \tre,
to $W(A^{g^{-1}})$, which is trivially gauge invariant under the extended gauge 
transformation
$$
\eqalign{
&A\to A^h,~~~X\to X^h\cr
&g\to hg\cr}
\eqn\quindici
$$
and thus the model is invariant up to one loop level.

We have repeated here the above well known manipulations [7] to emphasize the 
relevance of the invariance conditions 1) and 2) (eq.s \nove\ and \undici).
 
In many familiar example, such as the chiral Schwinger model, these conditions
are trivially satisfied. 

One well known case where these conditions 
become problematic is the $2d$ induced gravity or off-critical string. 
In this case, if one fixes 
the path integral measures ${\cal D} \f$ for the Weyl 
factor of metric and ${\cal D}\s$ for the Weyl group element 
by the invariance under the diffeomorphisms of $2d$ manifold,
then they are not invariant under the translations, eg. $\s\to\s+\a$ (i.e.
Weyl transformation). Thus, the path integral measure (i.e. ${\cal D} A
{\cal D} g$)
can never be invariant under the whole gauge group
$$
G=Diffeo\otimes Weyl
$$

\noindent{\it b) BRST [10] quantization}

A more rigorous strategy to have a consistent formulation of a 
gauge field theory is to recast
it in the BRST formalism. In this way, one may discuss the 
physically important questions such as the unitarity of S-matrix [11].

In the simpler example like the chiral gauge field theory where the 
invariance of the measure 
${\cal D} g {\cal D} A$, eq.s \nove\ and \undici, 
under the gauge transformations are respected,
there is no difficulty in setting up the BRST procedure once the anomaly 
has been removed.

One replaces the "heuristic" FP factor
$$
\D(A)\delta(F(A))=det \left ( {\delta F(A^h)\over \delta h}\Big|_{h=1}
\right) \delta (F(A))
$$
with BRST gauge fixing term
$$
exp-\int\hat s (\bar c F(A))= exp-\int\Big[B F(A)-\bar c {\delta F(A^h)\over
\delta h}\Big|_{h=1} c\Big]
$$
where $c,\bar c$ are the BRST ghosts corresponding to the gauge group $G$
while $B$ ("Lagrange multiplier") is the Nakanishi-Lautrup field. 
Under the BRST operator $\hat s$, one has, in particular
$$
\eqalign{&\hat s \bar c= B\cr
&\hat s B = 0\cr
&(\hat s^2=0)\cr}
$$

The counter term $\a(A;g)$ cancelling the one-loop anomaly, one can show 
easily the validity of the Slavnov-Taylor identity
$$
\eqalign{
&{\dd \tilde\G\over\dd A}{\dd \tilde\G\over\dd K}+{\dd \tilde\G\over\dd \F_i}
{\dd \tilde\G\over\dd K_i}+{\dd \tilde\G\over\dd c}{\dd \tilde\G\over\dd L}\cr
&(\tilde\G*\tilde\G=0)\cr}
\eqn\sedici
$$
up to one loop.

$\tilde\G$ is the generating functional of the one particle irreducible
part $\G$ (with added external source for composite operators) minus the 
"gauge fixing term" (in \sedici, $A$ and $c$ are the classical counter
parts of the gauge fields $A$ and ghost $c$, while $\{\F_i\}$ are the classical
fields for the matter $X$ and newly introduced field $g$; $K,K_i$ and $L$ are 
the usual external sources for the gauge variations $\hat\dd A,\hat\dd\F_i$
and $\dd c$ respectively).
One then hopes that it is possible to chose the higher order local counter
term in such a way that eq. \sedici\ is satisfied to all orders.

Let us now imagine, however, that the invariance conditions 1) and 2)
for the measure ${\cal D} A {\cal D} g$ 
(eq.s \nove\ and \undici) are not satisfied [11A].
This means that one should take account of one or both of the following
situations:

(1') the condition (1) is not satisfied, i.e. ${\cal D} A\not={\cal D} A^g
={\cal D} A e^{-\a'(A;g)}$, where $\a'(A;g)$ is the "Fujikawa determinant"
associated with the non gauge invariance of measure over gauge field
itself.

(2') the condition (2) is not satisfied, i.e. $\D(A^g)\not=\D(A)$.

First of all, the non invariance property 2') means that the factor
$\D (A)\dd (F(A))$ in eq. \undici\ must be replaced by $\D(A^{g^{-1}})
\dd(F(A))$. 

Thus, instead of a BRST gauge fixing term \quattordici\ one ends up with 
$$
\int \hat s (\bar c F(A)+ln\left( {\D(A^{g^{-1}})\over \D (A)}
\right)
\eqn\diciannove
$$
The trouble is that one can not transform $-ln\D (A)$ into a BRST invariant
local term in the action. In fact, the BRST gauge fixed action would appear
something like
$$
S_{\picc eff}=S_{\picc 0}+\a(A;g^{-1})+\a'(A;g^{-1})+
ln\left( {\D(A^{g^{-1}})\over \D (A)}\right)+\int \hat s (\bar c F(A)
\eqn\venti
$$

The extra one loop term $\a'(A;g)$ does not cause any trouble for the 
BRST scheme to work
at least in the example we are interested. One way to push through the
BRST scheme may be to replace eq. \venti\ with
$$
S_{\picc eff}'=S_{\picc 0}+\a(A;g^{-1})+\a'(A;g^{-1})+\int \hat s (\bar c F(A)
\eqn\ventuno
$$

It is likely that the effective action \ventuno\ leads to a consistent BRST
quantization. One may only add that it does not correspond to the
path integral method of ref.s [7] and [8] when $\D(A^g)\not=\D(A)$.

To reconciliate the "path integral" formulation of FS method with BRST
scheme, we propose another possibility.

It must be realized that once the new degree of freedom $g$ is admitted in the 
theory then there is no reason to exclude new local counter terms of the
right dimension which are BRST invariant and which may also depend
on $g$. Naturally this will 
change the model and its "physics", but nevertheless
it can remain consistent, in so far as the BRST invariance is maintained.

Let us then introduce the following counter term in our theory
$$
\tilde\L_{\picc G}(A,g;c,\bar c,c',\bar c',B)=\Big[ B G(A^{g^{-1}})-
\bar c'{\delta G(A^{g^{-1}h})\over \delta h}\Big|_{h=1} c'\Big]-
\Big[ B G(A)-
\bar c{\delta G(A^{h})\over \delta h}\Big|_{h=1}c \Big]
\eqn\ventidue
$$
where the second pair of "ghosts" $c',\bar c'$ are defined as the BRST singlet
$$
\eqalign{&\hat\delta\bar c'=0\cr
&\hat s c'=0\cr}
\eqn\ventitre
$$
and $G(A)$ is the "pseudo gauge fixing" which is generally different 
from $F(A)$.

The first term in $\tilde \L_{\picc G}$ is trivially BRST invariant since all 
the fields involved are either gauge invariant by themselves 
or appear as invariant 
combinations.
The second term, on the other hand, can be written as
$$
\hat s(\bar c G(A))
$$
so it is invariant too.

The effective action now takes the form
$$
S_{\picc eff}=S_{\picc 0}+\a(A;g^{-1})+\a'(A;g^{-1})+
\int\tilde\L_{\picc G}(A,g;c,\bar c,c',\bar c', B)+\int \hat s (\bar c F(A)
\eqn\ventiquattro
$$

Note that the gauge freedom of the BRST invariant theory \ventiquattro\
is represented by the (arbitrary) gauge fixing 
function $F(A)$ while each different 
choice of "pseudo gauge function" $G(A)$ defines
a new model.

Each choice of $G(A)$ then results in a gauge
invariant model which must then be gauge fixed by choosing a particular form
for $F(A)$.
In the limit of singular gauge
$$
F(A)\to G(A)
\eqn\venticinque
$$
the effective action \ventiquattro\ gives the series of models depending
on $G(A)$ alone. The corresponding effective action can be formally written
$$
S_{\picc eff}=S_{\picc 0}+\a(A;g^{-1})+
\a'(A;g^{-1})+\Big [ B G(A^{^{-1}})-\bar c' 
{\delta G(A^{g^{-1}h})\over\delta h}\Big|_{h=1} c'\Big]
\eqn\ventisei
$$

Note that in \ventisei\ the gauge is already fixed (with a singular gauge). 
To see the gauge invariance property of the model \ventisei, one must go back 
to eq. \ventiquattro\ with \ventidue, i.e.
$$
\eqalign{
S_{\picc eff}^{\picc inv}&=S_{\picc 0}+\a(A;g^{-1})+
\a'(A;g^{-1})+\int\Big [ B G(A^{^{-1}})-\bar c' 
{\delta G(A^{g^{-1}h})\over\delta h}\Big|_{h=1} c'\Big]\cr
&-\int\Big [ B [F(A)-G(A)]-\bar c{\delta \over\delta h}[
F(A^h)-G(A^h)]\Big|_{h=1} c\Big]\cr}
\eqn\ventisette
$$

We have seen in this way that the FS method of formulating an anomalous theory
within the path integral formalism apparently generates a series of physically
distinct and BRST invariant gauge fields theories.

We will discuss the possible candidate for such a scenario in the next section.

\vglue0.4cm
{\bf 3. Two dimensional induced gravity}
\vglue0.6cm
In this section we would like to apply the FS method of \S 1 to analyze the 
quantization problem of $2d$ gravity [13] 
(off critical string) in conformal
gauge [14].
The theory at classical level is defined in term of the Polyakov action
$$
S_{\picc 0}=\sum_{\mu=1}^d\int d^2 x \sqrt{g}g^{ab} \d_a X_\mu \d_b X^\mu
\eqn\ventotto
$$
where $\{X^\mu(x)\}_{\mu=1,d}$ are the bosonic matter fields coupled to 
the $2d$ metric $g_{ab}$ (in the string language, 
the string is immersed in a $d$-dimensional target
space).

We use euclidian metric and introduce the complex coordinates
$$
\eqalign{
&z=x_1+ix_2\cr
&\z=x_1-ix_2\cr}
$$

The invariant line element can be written as 
$$
ds^2=g_{ab}dx^a d x^b= e^\f |dz+\mu d\z|^2
\eqn\ventinove
$$

Thus, one can conveniently parametrize the metric as
$$
\eqalign{
&g_{zz}=\bar\mu e^\f,~~~~g_{\z\z}=\mu e^\f\cr
&g_{z\z}=g_{\z z}={1+\bar\mu\mu\over 2} e^\f\cr}
$$

In term of the parameters $\mu,\bar\mu$ and $\f$ the classical action
\ventotto\ takes the form [15]
$$
S_{\picc 0}=\sum_{\mu=1}^d\int {dz\wedge d\z\over 2i} {(\bar\d-\mu\d)X_\mu 
(\d-\bar\mu\bar\d)X^\mu\over 1-\mu\bar\mu}
$$

It is understood that $\mu$ and $\bar\mu$ are constrained by 
$$
|\mu|^2<1
$$

The classical action $S_{\picc 0}$ is invariant under the gauge
group $G$ which is the semidirect product of Diffeomorphisms
(general coordinate transformations)
and Weyl transformations. These symmetry groups imply respectively:

\noindent
1) the symmetry under the general coordinate transformation
$$
\eqalign{
&z\to z'=f(z,\z)\cr
&\z\to\z'=\bar f(z,\z)\cr}
\eqn\trenta
$$
where the relevant fields transform as follows
$$
\eqalign{
&X^\mu(z,\z)\to X^\mu\prime(z',\z')=X^\mu(z,\z)~~~(scalar)\cr
&\mu(z,\z)\to\mu'(\z',\z')=-{\bar\d f-\mu\d f\over\bar\d\bar f-\mu\d \bar f}
(z,\z)\cr
&\f(z,\z)\to\f'(z',\z')=\f(z,\z)+
ln 
{(\bar\d \bar f-\mu\d \bar f)(\d f-\bar\mu \bar\d f)\over D_{f}^{2} }\cr}
\eqn\trentuno
$$
where 
$$
D_f=det\left(\matrix{&\d f&\d\bar f\cr &\bar \d f&\bar\d\bar f\cr}\right)
$$

\noindent
2) The symmetry under the local rescaling of the $2d$ metric
$$
g_{ab}\to e^\s g_{ab}
$$
or in term of the $\mu,\bar\mu$ and $\f$ variables
$$
\mu\to\mu,~~~~\bar\mu\to\bar\mu,~~~~\f\to\f+\s
\eqn\trentadue
$$

It is well known that the theory is anomalous, i.e. one can not regularize the
path integral in a way that conserves the whole $G=Diffeo\times
Weyl$ group.

One can see this easily, examining the matter integral measure ${\cal D}X^\mu$.
With the simplest (translationally invariant or "flat") regularization
${\cal D}_{\picc 0}X^\mu$, one has
$$
\prod_{\mu=1}^d\int {\cal D}_{\picc 0} X^\mu e^{S_{\picc 0}(X,\mu,\bar\mu)}
=exp-{d\over 24\pi}[W(\mu)+\bar W(\bar\mu)]
\eqn\trentatre
$$
where $W(\mu)$ is the Polyakov's "light cone gauge" action [13].

This is naturally Weyl invariant ($S_{\picc 0}$ does not contain the variable
$\f$). On the other hand, it is equally clear that one has lost
diffeomorphism's invariance, since the invariance under general 
coordinate transformations means
$$
\dd W(\mu)=0
\eqn\trentaquattro
$$
under $\dd\mu=(\bar\d-\mu\d+\d\mu)(\e+\mu\bar\e)$, which corresponds to
the infinitesimal version of eq.s \trentuno\ with 
$f(z,\z)=\e(z,\z),\bar f(z,\z)=\bar\e(z,\z)$.

Eq. \trentaquattro\ is equivalent to the functional differential equation
$$
(\bar\d-\mu\d-2\d\mu){\dd W\over \dd\mu(z,\z)}=0
$$

A well known computation [16] gives, instead,
$$
(\bar\d-\mu\d-2\d\mu){\dd W\over \dd\mu(z,\z)}=\d^3\mu\not= 0
\eqn\trentacinque
$$

Thus, ${\cal D}_{\picc 0} X^\mu$ can not be invariant under diffeomorphisms.
One can define the diffeomorphisms invariant measure 
${\cal D}_{\picc Diffeo} X^\mu$ by introducing the local counter term 
$$
\eqalign{
\L(\mu,\bar\mu,\f)&=-{1\over 2}\int {dz\wedge d\z\over 2\pi} \Big[
{1\over 1-\mu\bar\mu}[(\d-\bar\mu\bar\d)\f(\bar\d-\mu\d)\f\cr
&-2(\bar\d\bar\mu(\bar\d-\mu\d)+\d\mu(\d-\bar\mu\bar\d))\f ]
+F(\mu,\bar\mu)\Big ]\cr}
\eqn\trentasei
$$
where $F(\mu,\bar\mu)$ is a local function of $\mu$ and $\bar\mu$ 
only. We do not need the explicit form of $F$ [17].

The new effective action 
$$
W_{\picc cov}(\mu,\bar\mu,\f)=W(\mu)+\bar W(\bar\mu)+\L(\mu,\bar\mu,\f)
$$
is invariant under diffeomorphisms.
 
One can write $W_{\picc cov}(\mu,\bar\mu,\f)$ compactly in the form
$$
W_{\picc cov}(\mu,\bar\mu,\f)=\int {dz\wedge d\z\over 2\pi}
{(\d-\bar\mu\bar\d)\F (\bar\d-\mu\d)\F\over 1-\mu\bar\mu}=
\int d^2 x \sqrt{g} g^{ab}\d_a\F\d_b\F
\eqn\trentasette
$$
where $\F=\f-ln\d\ze\bar\d\ze$ and $\mu={\bar\d\ze\over\d\ze}$ 
(Beltrami differentials).
Non local (with respect to $\mu$ and $\bar\mu$) parameter $\ze(z,\z)$
is Polyakov meson field (13) in $2d$ gravity. 

One characterizes the diffeomorphisms invariant measure ${\cal D}_{\picc diffeo}
X^\mu$ by
$$
\prod_{\mu=1}^d \int {\cal D}_{\picc diffeo}X^\mu e^{S_{\picc 0}(X,\mu,\bar\mu)}
=exp-{d\over 24\pi} W_{\picc cov}(\mu,\bar\mu,\f)
\eqn\trentotto
$$

(One can understand the appearance of $\f$ field, which is absent in the
classical action, as due to the introduction of a 
covariant regularization: 
$\L_{\picc cov}$, $ds^2\sim e^\f|dz|^2>\L^2_{\picc cov}$).

Following for instance DDK [14], in what follows we consistently make
use of the diffeomorphisms invariant measure. 
Thus, except when indicated explicitly otherwise, 
$$
{\cal D} X^\mu\equiv {\cal D}_{\picc Diffeo} X^\mu
\eqn\trentanove
$$
and more generally ${\cal D} \v\equiv {\cal D}_{\picc Diffeo}\v$ for any other
filed $\v$.

Evidently, the diffeomorphisms invariant measure ${\cal D} X^\mu$ can not
be invariant under the Weyl transformation
$$
\f\to\f+\s
$$

Thus, one establishes that the theory is $G$ anomalous.

(Faddeev-Shatashvili method)

Having seen that our model for $2d$ gravity is anomalous, one would like to
apply to it 
the FS method of "gauge invariant" quantization of \S 1. As in \S 1,
we "preestablish" the gauge choice for the full group $G=Diffeo\times Weyl$
$$
\eqalign{
&\mu=\mu_0\cr
&\bar\mu=\bar\mu_0~~~diffeomorphisms\cr
&F(\f)=0~~~~Weyl\cr}
\eqn\quaranta
$$

Since our regularization preserves the diffeomorphisms we assume that 
the gauge fixing problem (with relevant "$b,c$" ghosts) for diffeomorphisms
has been already taken care for.

To deal with anomalous Weyl symmetry, we have to introduce an extra 
degree of freedom, a scalar field $\s(z,\z)$, corresponding to the element 
of Weyl symmetry group $g=e^{\s(z,\z)}$.

The anomaly cancelling counter term suggested by FS is then given by
$$
\eqalign{
\a(\mu,\bar\mu,\f;-\s)&=W_{\picc cov}((\mu,\bar\mu,\f-\s)-W_{\picc cov}
(\mu,\bar\mu,\f)=\cr
&=-{1\over 2}\int {dz\wedge d\z\over 2i} {1\over 1-\mu\bar\mu}
[(\d-\bar\mu\bar\d)\s(\bar\d-\mu\d)\s+2(\d-\bar\mu\bar\d)\s(\bar\d-\mu\d)\f\cr
&-2(\bar\d\bar\mu(\bar\d-\mu\d)+\d\mu(\d-\bar\mu\bar\d))\f]\cr}
\eqn\quarantuno
$$

Note that the non local part of $W_{\picc cov}$ is cancelled and 
$\a(\mu,\bar\mu,\f;-\s)$ is perfectly local. Naturally, one needs the counter 
term $\a$ for each covariant one loop integral corresponding not only to the
matter field $\{X^\mu\}_{\mu=1}^d$, but also to the diffeomorphism ghosts, 
$b,c$ and $\bar b,\bar c$, as well as to the $\f$ field contained in 
$W_{\picc cov}(\mu,\bar\mu,\f)$.

Thus, the effective action in sense of \S 2 is given by
$$
S_{\picc eff}=S_{\picc 0}(X, \mu,\bar\mu)+S_{\picc gf}^{(d)}(b,c,\bar b,\bar c,
B,\bar B, \mu,\bar \mu)+\g'\a(\mu,\bar\mu,\f;-\s)
\eqn\quarantadue
$$
where $S_{\picc gf}^{(d)}$ is the gauge fixing term with respect to the
non anomalous diffeomorphism symmetry.

As explained above, the coefficient $\g'$ is contributed by all the relevant
fields, that is 
$\{X^\mu\}_{\mu=1}^d\Rightarrow d, (b,c,\bar b,\bar c)\Rightarrow
-26, \f\Rightarrow 1$, which gives $\g'={d-26+1\over 24\pi}={d-25\over 24\pi}$.

Note that the contribution of $\f$ field is due to the fact that 
${\cal D}_{\picc Diffeo}\f\not= {\cal D}_{\picc 0}\f$, or
in the terminology of \S 2, that one needs the "second" FS counter term
"$\a'(\f;\s)$". 

One can now write down the partition function $Z$ with the FS prescription
(within the path integral formalism of ref. [7], see eq. \otto\ of
\S 1). Integrating out the "matter fields" ($X^\mu,b,c,\bar b,\bar c, $),
one has
$$
\eqalign{
Z\sim&\int{\cal D}\s{\cal D}\f \Big [ exp -\g'\int {dz\wedge d\z\over 2i}
{1\over 1-\mu_0\bar\mu_0}\big(
(\d-\bar\mu_0\bar\d)(\f-\s)(\bar\d-\mu_0\d)(\f-\s)\cr
&-2(\bar\d\bar\mu_0(\bar\d-\mu_0\d)+\d\mu_0(\d-\bar\mu_0\bar\d))(\f-\s)
\big)\Big]\D(\f-\s)\dd(F(\f))\cr}
\eqn\quarantatre
$$
where the local action in the exponential is essentially a Liouville
action $S_{\picc L}'(\f')$, ($\f'=\f-\s$). The last two factors come from
the $\dd$ function insertion
$$
\D(\f)\int{\cal D}\s\dd(F(\f+\s))=1
\eqn\quarantaquattro
$$

Note that, since ${\cal D}\s\equiv{\cal D}_{\picc Diffeo}\s\not=
{\cal D}_{\picc 0}\s$ (${\cal D}_{\picc 0}\s$ "flat" measure)
$$
\D(\f-\s)\not=\D(\f)
\eqn\quarantacinque
$$

Formally, one can write the $\D(\f-\s)$ factor as a local action with
the help of the "Weyl ghosts" $\psi$ and $\bar\psi$
$$
\D(\f-\s)=\int{\cal D}\psi{\cal D}\bar\psi exp-\int \bar\psi{\dd F(\f-\s)
\over\dd\f}\psi
\eqn\quarantasei
$$

(BRST procedure)

The path integral argument of \S 1 is at best heuristic. It may suggest the
possible models but one can not prove in this way their consistency.

As it has been argued in \S 1, one may start a more precise discussion
after setting up the BRST quantization procedure. The BRST properties of 
the type of models we are dealing with here, have been studied in details
for the critical case, i.e. for $d=26$, where the theory is not
anomalous. In ref. [15], the BRST transformation properties of the fields 
are given. They may be used to study our (off critical) model.

One has (see eq. \trentuno)
$$
\eqalign{
&\hat\dd X^\mu=(\xi\cdot\d)X^\mu\cr
&\hat\dd \mu=(\bar\d-\mu\d+\d\mu)c\cr
&\hat\dd\f=\psi+(\xi\d)\f+(\d\xi)+\mu\d\bar\xi+\bar\mu\bar\d\xi\cr
&\hat\dd\xi=(\xi\cdot\d)\xi\cr
&\hat\dd c= c\d c\cr
&\hat\dd\psi=(\xi\cdot\d)\psi\cr}
\eqn\quarantasette
$$
where $\xi\cdot\d$ means $\xi\d+\bar\xi\bar\d$.

Here $\hat\dd$ stands for the both Weyl and diffeomorphism symmetries.
The diffeomorphism ghosts $c,\bar c$ are related to the original 
($\xi,\bar\xi$) (corresponding to $\dd z=\e(z,\z),\dd\z=\bar\e(z,\z)$)
by
$$
\eqalign{
&c=\xi+\mu\bar\xi\cr
&\bar c=\bar\xi+\bar\mu\xi\cr}
\eqn\quarantotto
$$

To eq. \quarantasette, we must add the transformation of the auxiliary
field $\s(z,\z)$. Since $\s$ must be a scalar with respect to diffeomorphisms
one has
$$
\hat\dd\s=\psi+(\xi\cdot\d)\s
\eqn\quarantanove
$$

Together with the formulae in eq.s \quarantasette\ to \quarantanove, 
one consistently finds
$$
\hat\dd^2=0
\eqn\cinquanta
$$

One should add also the diffeomorphisms anti ghost ($b,\bar b$) and Weyl 
anti ghost $\bar\psi$ with the corresponding Nakanishi-Lantrup fields
$B$ and $D$. Their transformation properties are
$$
\eqalign{
&\hat sb=B,~\hat s \bar b=\bar B,~\hat s\bar\psi= D\cr
&\hat s B=\hat s \bar B=\hat s D=0\cr}
\eqn\cinquantuno
$$

We have seen, however, that the Faddeev-Popov factor $\D(\f)$ is not Weyl 
invariant \quarantacinque. Thus, according to the result of \S 1, one needs
to correct the effective action $S_{\picc eff}$ by modifying the factor
$\D(\f-\s)\dd(F(\f))$ into a BRST gauge fixing term. As we have seen in \S 1,
such a prescription is not unique. Formally, any action of the form
$$
BRST(invariant)+\hat s(\psi F(\f))(BRST exact)
$$
will do the job.

Now the factor $\D(\f-\s)\dd(F(\f))$ can be rewritten in the form
$$
exp-\int\Big( {\cal D}F(\f)+\bar\psi'{\dd F\over\dd \f}(\f-\s)\psi'\Big)
$$

Thus, in order to follow this expression as close as possible, we suggest
to add a counter term of the form of eq. \ventidue\ in \S 1
$$
\tilde\L_{\picc G}(\f,\s;\psi,\bar \psi,\psi',\bar \psi',D)=
\Big[ D G(\f-\s)+\bar\psi'{\dd G\over\dd\f}(\f-\s)\psi'\Big]-
\Big[DG(\f)+\bar\psi{\dd G\over\dd\f}(\f)\psi\Big]
\eqn\cinquantadue
$$
where we have introduced the function $G(\f)$ to distinguish it from
the true gauge fixing term $s(\bar\psi F(\f))$. The new fields $\psi'$ and
$\bar\psi'$ in eq. \cinquantadue\ ($c'$ and $\bar c'$ in eq. \ventidue\ )
are Weyl singlet and transform as
$$
\eqalign{
&\hat\dd\bar\psi'=0\cr 
&\hat\dd\psi'=(\xi\cdot\d)\psi'\cr}
\eqn\cinquantatre
$$

With the addition of the counter term $\tilde\L_{\picc G}$, the effective 
action now reads
$$
\eqalign{
\tilde S_{\picc eff}&=S_{\picc L}''(\f-\s)+\int \tilde\L_{\picc G}(\f,\s;\psi,
\bar\psi,\psi',\bar\psi',D)+\int \hat s (\bar\psi F(\f))\cr
&=S_{\picc L}''(\f-\s)+
\int\Big[DG(\f-\s)+\bar\psi'{\dd G\over\dd\f}(\f-\s)\psi'\Big]
+\int\hat s(\bar\psi(F-G)(\f))\cr}
\eqn\cinquantaquattro
$$

The expression for $\tilde S_{\picc eff}$ contains two arbitrary
functions $F(\f)$ and $G(\f)$. Their roles are completely different. While
$F(\f))$ is a genuine gauge fixing function, each choice of $G(\f)$ 
actually defines a new model.    

Naturally, the "series" of models (at arbitrary gauge) includes the familiar
cases. For example, if one fix the model by choosing
$$
G=0
$$
one reproduces the physically equivalent formulations of DDK model.

Alternatively, for any given $G$, one may consider the singular gauge limit 
$$
F\to G
$$

In this limit the model formally corresponds to the action
$$
\tilde S_{\picc eff}=S_{\picc L}''(\f-\s)+
\int\Big[DG(\f-\s)+\bar\psi'{\dd G\over\dd\f}(\f-\s)\psi'\Big]
\eqn\cinquantasette
$$

This is the type of model treated in ref. [18]. One may further add the 
BRST invariant term $-{\l\over 2}\int D^2$ and transform $\tilde S_{\picc eff}$
into
$$
\tilde S_{\picc eff}'=S_{\picc L}''(\f-\s)+
\int\Big[{1\over 2\l}G^2(\f-\s)+\bar\psi'{\dd G\over\dd\f}(\f-\s)\psi'\Big]
\eqn\cinquantotto
$$

Eq. \cinquantasette\ (or \cinquantotto) seems to be the closest BRST
quantizable approximation to the consequence of FS prescription, i.e. the
insertion
$$
1=\D(\f)\int{\cal D}\s \dd(G(\f+\s))
\eqn\cinquantanove
$$

In ref. [18], and in some later works, the choice
$$
G(\f)=R(\f)-R_0
\eqn\sessanta
$$
with $R$ the scalar curvature, has been made. Using \sessanta, the effective
action \cinquantotto\ becomes
$$
\tilde S_{\picc eff}'((\f'=\f-\s),\psi',\bar\psi,)=
S_{\picc L}''(\f')+
\int\Big[{1\over 2\l}(R(\f')-R_0)^2(\f-\s)+
\bar\psi'{\dd R\over\dd\f}(\f-\s)\psi'\Big]
\eqn\sessantuno
$$

Note that the model defined by \sessantuno\ is fully interacting. In particular
{\it a}) the presence of propagating $\psi'$ and $\bar\psi'$ fields and
{\it b}), more importantly, the presence of $\psi', \bar\psi'$ and 
$\f'$ (Yukawa)
interaction in \sessantuno, change the parameters in the Liouville type action
$S_{\picc L}''(\f')$. Such a change, which affects the low energy dynamics
of \sessantuno, can not be calculated exactly. It is not easy even to develop
a systematic perturbation expansion [20]. 
We believe [18] [19] that the modification
represented by eq. \sessantuno\ may result in deviations from the classical
DDK result, when one uses \sessantuno\ to calculate such physical quantities
as string tension and anomalous dimension.

Lastly, it must be mentioned that the BRST invariant term
$$
\int\bar\psi'{\dd G\over\dd\f}(\f-\s)\psi'
\eqn\sessantadue
$$
in \cinquantasette\ could also be obtained from the alternative gauge fixing
$$
S_{\picc gf}=\int \hat s\Big [\bar\psi{\dd G\over\dd\f}(\f-\s)\s\Big]
\eqn\sessantatre
$$

In this case, one can dispense with the extra BRST invariant (for Weyl 
transformation) $\psi'$ and $\bar\psi'$ degrees of freedom. The gauge 
fixing function is
$$
F(\f,\s)={\dd G(\f-\s)\over\dd\f}(\f-\s)\s
\eqn\sessantaquattro
$$

It looks as if this model is gauge equivalent
to the DDK model, since the gauge choice $G(\f)=\f$ gives the effective 
action
$$
S_{\picc eff}=S_{\picc L}'(\f-\s)+\int(\psi\bar\psi+D\s)\sim S_{\picc L}'(\f)
~~~~(\s=0)
\eqn\sessantacinque
$$

The Liouville action $S_{\picc L}'$ here is identical to eq. \quarantaquattro\
without further renormalization (eq. \sessantuno\ is a free field action).

In the next section, we apply the DDK [14] type consistency arguments to
analyze the consequences of the model (eq. \sessantuno), paying attention
to the influence of the Yukawa term $(\bar\psi'\psi' \phi)$ in \sessantuno.

\vglue0.4cm
{\bf 4. Physical consequences (modified KPZ-DDK model)}
\vglue0.6cm

After reading the last section, one may wonder if the
counter term such as
$$
DG(\f-\s)+\bar\psi'{\delta G\over\delta\phi}(\f-\s)\psi'
\eqn\nuovauno
$$
(eq. \cinquantadue\ of \S 3) may indeed influence the physics in any way.
in fact, it is very probable that such an influence is washed away for large
class of "pseudo gauge functions" $G(\f)$ by the renormalization group argument.

However, for the specific choice of ref. [18], i.e. (eq. \sessanta\ of \S 3)
$$
G(\f)=R(\f)-R_0
$$
it gives actually the possibility of modifying the classical KPZ results
on the string tension and anomalous conformal dimensions.

The effective action which corresponds to the above choice of $G(\f)$ is
given by eq. \sessantuno. As it has been remarked previously, this action 
is equivalent to the well known Kawai-Nakayama $R^2$ model [21], if one 
omits precisely the "fake" FP term
$$
\int\bar\psi'{\dd R\over\dd\f}(\f-\s)\psi'
\eqn\nuovadue
$$
in eq. \sessantuno.

Now, in ref. [21], it has been shown  that the Kawai-Nakayama model with 
$R^2$ (or $(R-R_0)^2$) term gives the same scaling behaviour for large 
distance as the original DDK model. That is, for the fixed area
partition function 
$$
Z(A)=\int{\cal D} (fields)(Jacobians) exp-S_{\picc eff}\times \delta
\left(\int dx^2\sqrt{g}-A\right)\sim const~A^{-\G(h)-3}
\eqn\nuovatre
$$
as $A\to\infty$, except for ${1\over A}$ correction in the exponent.
The string tension is the same as the KPZ result ($h$ is the genus)
$$
\G(h)=(1-h) {25-d+\sqrt{(1-d)(25-d)}\over 12}
\eqn\nuovaquattro
$$

It is not a simple matter to calculate the possible change with respect to
this result in the presence of pseudo FP term \nuovadue.
The difficulty is due to the fact that we have here the genuine interacting
theory instead of effective gaussian model such as the original DDK case [14].

Here we present the approximate analysis which is, at best, valid for
the low energy (large distance) regime.

Writing down the pseudo FP term \nuovadue\ in detail, we have
$$
S''=\int {dz\wedge d\z\over 2i} \sqrt{\hat g}\bar\psi'(-\d\bar\d+\d\bar\d
(\f-\s)+\hat R)\psi'
\eqn\nuovacinque
$$
(in the conformal gauge where $\mu=\mu_0=0, \bar\mu=\bar\mu_0=0$).

In $S''$, the free part for $\bar\psi'$ and $\psi'$ has the structure of the
so called $bc$ ghost system
$$
S''(free)=\int b\bar\d c {dz\wedge d\z\over 2i}
\eqn\nuovasei
$$
if one identifies
$$
\eqalign{
&b\equiv\d\bar\psi'\cr
&c\equiv\psi'\cr}
\eqn\nuovasette
$$
with the stress energy tensor and ghost number current given by
$$
\eqalign{
&T=-b\d c\cr
&J=bc\cr}
\eqn\nuovaotto
$$
(note that the conformal dimension of $b$ and $c$ here are respectively 1 and
0).

Then one can give an equivalent bosonic system with
$$
\eqalign{
&T'=-{1\over 2} ((\d\v)^2+Q'\d^2\v)\cr
&J'=i\d\v\cr}
\eqn\nuovanove
$$
which reproduces the same algebraic structure as the system \nuovaotto\ 
for the suitable value of $Q'(=i)$, if the new scalar field $\v$ satisfies
$$
\v(z)\v(w)\sim-\log |z-w|^2
$$

One should still take account of the interaction (Yukawa) term in \nuovacinque.
To do so, we write
$$
\d\bar\psi'\psi'=iA\d\v+...
\eqn\nuovadieci
$$
where the $...$ represent the higher order corrections. 

Thus, the low energy equivalent of \nuovacinque\ is
$$
S''\sim\int \sqrt{\hat g} (-\v\d\bar\d\v-i(1+A)\v \hat R+i\a A\v\d\bar\d\f')
\eqn\nuovaundici
$$
($\f'=\f-\s$, see \S 3).

The undetermined constant $A$ represents the first order correction due
to the interaction. The constant $\a$ is the usual gravitational correction
($g_{ab}=e^\f\hat g_{ab}\to e^{\a\f}\hat g_{ab}$).

To avoid the 
imaginary coupling constant in \nuovaundici\ (problem of unitarity in BRST
approach), we "Wick rotate" $\v$, $\v\to i\v$.

Then, with the redefinition of constants in \nuovaundici, one can write
a low energy approximation as
$$
S''={1\over 8\pi} \int \sqrt{\hat g} (\v\d\bar\d\v-2B\v\d\bar\d\f' 
+\tilde Q\v \hat R)
\eqn\nuovadodici
$$

Putting this together with the rest of the effective action in eq. \sessantuno,
our low energy approximation consists of taking a gaussian model with two 
scalars
$$
\tilde S'_{\picc eff}(\f',\v)={1\over 8\pi}\int {dz\wedge d\z\over 2i}
\sqrt{\hat g}(-M_{ij}\F^i\qgg \F^j - Q_i\hat R\F^i)
\eqn\nuovatredici
$$
where
$$
\eqalign{
&\F^i\equiv (\F^1,\F^2)=(\f',\v)\cr
&Q_i \equiv (Q_1,Q_2)=(Q,- \tilde Q)\cr}
$$
and
$$
M_{ij}=\left(\matrix{1&B\cr B&-1\cr}\right)
$$

Such a 2-bosonic system with "Lorentzian" metric as a kind of improvement
over the standard Liouville type 2$d$ gravity (DDK model) has been suggested
in the past [18] [23]. More recently, Cangemi, Jackiw and Zwiebach have given
the thorough field theoretical analysis of such a system for $B=0$, treating
it as the "dilatation gravity" [24]. 
Here, however, the presence of $\v-\f'$ coupling term ($B\not=0$) is crucial
for the possible modification of KPZ-DDK result.

The origin of such a term is, of course, the Yukawa coupling in the original
counter term \nuovauno.

At this point, one can in principle apply the technics of ref. [20] to get
the perturbative estimate of the constant $A$ (i.e. $B$). We leave such an 
analysis for further publication and content ourselves with repeating the 
original DDK consistency arguments to indicate that indeed one has the 
possibility of changing the KPZ-DDK result.

Thus, we would like to apply the effective action \nuovatredici\ to estimate

\noindent {\it a}) string tension $\G(h)$, and

\noindent {\it b}) renormalization $\D_0\to\D$ of the conformal dimension
of a primary operator ${\cal O}$.

From the effective action \nuovatredici, one can derive the expression for the
gravitational stress energy tensor
$$
T_{\picc grav}=-{1\over 2} (M_{ij}\d\F^i\d\F^j+B Q_i \d^2 \F^i)
\eqn\nuovaquattordici
$$
which contributes to the central charge by the amount
$$
c_{\picc grav}=2+3 M^{ij}Q_i Q_j
\eqn\nuovaquindici
$$
where $M^{ik}M_{kj}=\delta^i_j $.

\noindent {\it a) String tension}

The detail of how to generalize DDK argument to get the string tension 
$\G(h)$ in our model is given in Ref. [18]. We limit ourselves, therefore,
to give the more relevant results.

The consistency conditions lead to the determination of $Q_i$'s as
$$
\eqalign{
&Q_1= -{1\over \sqrt{3}} \left(B\sqrt{1+Bd}-\sqrt{1+B^2+(B-1)d}\right)\cr
&Q_2=\sqrt{{1+Bd\over 3}}\cr}
\eqn\nuovasedici
$$

Then the string tension $\G(h)$ is given, just as in ref. [14], by
$$
\G(h)=\chi(h){Q_1\over\a}+2
$$
($\chi(h)=2(1-h)$ is the Euler index).
$\a$ can be calculated again as in ref. [14] from 
$$
dim(e^{\a\f'}\sqrt{\hat g})=1
$$
which gives
$$
\a=-{\sqrt{1+B^2}\over 2\sqrt{3}}\left(\sqrt{25+(B-1)d}-\sqrt{1+(B-1)d}\right)
\eqn\nuovadiciassette
$$

Thus, one obtains the string tension in our model as the function of $B$
$$
\G(h)={2(1-h)\over\sqrt{1+B^2}} {B\sqrt{1+Bd}-\sqrt{(1+B^2)(25+(B-1)d}
\over \sqrt{25+(b-1)d}-\sqrt{1+(B-1)d}}+2
\eqn\nuovadiciotto
$$
which reduces to the KPZ expression if $B=0$ (i.e. $d\leq 1$). Note that
$\G(h)$ is real for $B\geq 1-{1\over d}$ with an arbitrary positive $d$.
In view of excellent agreement between KPZ formula and "experiments" for 
$d<1$, we might expect some sort of phase transition behaviour
$$
B\propto \t (d-1)
$$
(where $\t(x)$ is the step function) but it would be very hard to show such
a behaviour by the limited technics available [20]. For the various 
"improvements" and applications to statistical mechanic of \nuovadiciotto\
we refer to [19].

\noindent {\it b) Anomalous dimension}

The calculation of the renormalization of conformal dimension is more 
straight forward.

Let the base conformal dimensions of an operator ${\cal O}$ 
be $(\D_0, \bar\D_0)$.
We would like to construct the globally defined operators $\int e^{\a\f}
\sqrt{\hat g}$ and $\int e^{\b\f}\sqrt{\hat g}{\cal O}$. 
This requirement implies
$$
\eqalign{
&dim(e^{\a\f}\sqrt{\hat g})=(1,1)\cr
&dim(e^{\b\f}\sqrt{\hat g}{\cal O})=(1,1)\cr}
$$

These conditions give rise to the two equations
$$
\eqalign{
&\a^2+(Q_1+BQ_2)\a+2(1+B^2)=0\cr
&\b^2+(Q_1+BQ_2)\b+2(1+B^2)(1-\D_0)=0\cr}
\eqn\nuovadiciannove
$$

The renormalized dimension $\D$ of the operator ${\cal O}$ can be read off from
the asymptotic formula
$$
F_{\cal O}(A)=\int {\cal D}\f' {\cal D}\v e^{-\tilde S_{\picc eff}'}\delta
\left(\int e^{\a\f'}\sqrt{\hat g}-A\right)
\int {\cal O} e^{\b\f'}\sqrt{\hat g}\Big / Z(A)\sim K_{\cal O} A^{1-\D}
$$

This gives, just as in ref. [14], 
$$
1-\D={\b\over\a}
\eqn\nuovaventi
$$

From eq.s \nuovadiciannove\ and \nuovaventi, one gets the equation determining
$\D$
$$
\D-\D_0= -{1\over 2}{1\over 1+B^2} \a^2 \D(\D-1)
$$

Needless to say, this too reduces to KPZ result when $B=0$.

\vglue0.4cm
{\bf 5. Conclusion}
\vglue0.6cm

In this note, we have tried to analyze further consequences of the 
Faddeev-Shatshvili method of quantizing anomalous gauge fields theories.

In contrast with other authors [5A], we did not try to
show the equivalence with the "gauge non invariant" method of which
the Jackiw-Rajaraman treatment of the chiral Schwinger model is
 a distinguished example. On the contrary, we have argued that, in certain
cases of physical interest, the FS method can be used to generates new models.

The series of "new" $2d$ gravity models proposed here includes the 
models in ref.s [18] [19] as well as the Kawai-Nakayama type $(R-R_0)^2$
(or $R^2$) models [21] [22].

The analysis presented in \S 4 with respect to the model 
eq. \sessantuno\ is at best heuristic and we certainly cannot (and do not)
claim to have "solved" the famous $d=1$ barrier problem in 2-$d$ gravity.
We merely indicate possible ways to modify the original DDK model.

To see if the possibility of enlarging in this way the $2d$ (induced) gravity
models really throws some light on the problem of the $d=1$ barrier
in $2d$ gravity, we need a more thorough analysis of the consistency 
of these models as well as a better understanding of their physical 
consequences.

One would like to end by mentioning a further peculiarity about the anomalous
Dif\-fe\-o\-mor\-phism-Weyl gauge symmetry of 2$d$ gravity.

It is natural to ask whether, instead of somehow trying to conserve the
entire gauge symmetry of the anomalous classic model, one may
still have a physical consistent quantum model by keeping only the "maximal"
anomalous free part of the classical symmetry (up to local-counterterms).

Recently precisely such a suggestion has been made by R. Jackiw and others
[25].
They counter the conventional argument favouring the Diffeomorphism symmetry 
(over Weyl) by pointing out the even greater difficulty of conserving
the whole Diffeomorphism symmetry in the quantum canonical hamiltonian approach
[26].

Thus, in ref. [25], it has been suggested to conserve Weyl symmetry plus
area (volume) preserving diffeomorphim (i.e. Diffeomorphism $x^{\mu}\to
x^{\mu}\prime=f^{\mu} (x)$ with the 
constrain $\det \left[ {\d f^\mu\over\d x^\nu}\right] =1.$)

Jackiw's formalism can be generalized to the series of models which are 
symmetric under the modified Diffeomorphisms $D^{(k)}$:
$$
\eqalign
{&x^\mu \to x^{\mu}\prime=f_{(k)}^\mu (x)\cr
&g_{\mu\nu}(x)\to g_{\mu\nu}\prime(x')=g_{\alpha\beta}(x)
{\d x^\a \over \d x^\mu\prime}
{\d x^\b\over \d x^\nu\prime}\left[ \det\left({\d x^\eta\over\d 
x^{\lambda}\prime}\right)
\right]^{k-1\over k}, \cr}
\eqn\ultima
$$
where $0<k\leq 1$. While $k=1$ corresponds to the usual Diffeomorphism 
invariant DDK like model, the limit $k\to 0$ can be shown to give the 
improved Weyl invariant model of Jackiw et al.

Superficially, these models parameterized by $k$ corresponds to different
gauge symmetries and, in particular, one migth expect a drastic change
of the physics between the two limits $k=1$ (Diffeomorphisms) and
$k\to 0$ (Weyl and Area preserving diffeomorphisms).

However, there are reasons to believe that they actually correspond to 
the same physisc. 

(1) One can move formally from the "standard"
$D^{(1)}$ invariant model to the $D^{(k\not= 1)}$ defined throught
Eq. \ultima\ by a simple changing of variables. In terms of Beltrami
parameterization of the 2$d$ metric in section 3; this changing
of variables is given by
$$
\eqalign{
&\mu\to \mu^{(k)} =\mu\cr
&\bar\mu\to \bar\mu^{(k)}=\bar\mu\cr
&\phi\to\phi^{(k)}=k\phi+(1-k) \log \left({1\over 1-\mu\bar\mu}\right)
=\phi+(k-1)\log\sqrt{-g}.\cr}
\eqn\ultimissima
$$ 

At quantum level, Eq. \ultimissima\ amounts to different choice of local 
counter-term.

(2) In ref. [27], the two dimensional Hawking radiation has been calculated
using Jackiw's Weyl invariant model as well as the general $D^{(k)}$ invariant 
model. In either case, the result is identical with the standard 
($D^{(1)}$ invariant) model. This fact means that, at least the black-hole thermodinamics is
independent from the parameter $k$. 

If one conjectures from these facts that the choice of invariant gauge
group is in some sense irrilevant (at least for 2$d$ gravity),
the implication for the anomalous gauge field theory is not clear.

\vglue 0.6cm
{\bf Acknowledgment}
\vglue0.4cm
This work has been completed while one of the authors (KY) stayed at 
National Laboratory of High Energy Physics, Tsukuba, Japan (KEK). 
It is a pleasure to thank prof. H. Sugawara and M. Ishibashi for the 
hospitality. KY acknowledges stimulating discussions 
with many members of KEK, in particular: H. Kawai, S. Aoki, T. Yukawa and
M. Ishibashi. The constructive comments from K. Fujikawa, N. Nakazawa and 
K. Ogawa are also gratefully acknowledged. The authors thank G.C. Rossi
for the thorough reading of the manuscript. The work is partially supported
by INFN and Italian Minister of Science and University, MPI 40\%.

\vglue 0.6cm
{\bf REFERENCES}
\vglue 0.4cm
\parindent=0.pt

{[1] C. Bouchia, J. Iliopoulos and Ph. Mayer, Phys. Lett. B38 (1972) 519;}

{[2] D. Gross and R. Jackiw, Phys. Rev. D6 (1972) 477;}

{[3] M. Kato and K. Ogawa, Nucl. Phys. B212 (1983) 443;}

{[4] R. Jackiw and R. Rajaraman, Phys. Rev. Lett. 54 (1985) 1219;}

{[5] R. Rajaraman, Phys. Lett. B154 (1985) 305;}

{[5A] There are large number of works accumulated on this subject.
We list below a few of them which seem to be relevant for the present
discussion. The list is by no means complete however.}

{H.O. Girotti, H.J. Rothe and K.D. Rothe, Phys. Rev. D34 (1986) 592;}

{I.G. Halliday, E. Rabinovici, A. Schwimmer and M. Chanowitz, Nucl. Phys.
B268 (1986) 413;}

{D. Boyanovsky, Nucl. Phys. B294 (1987) 223;}

{L. Caneschi and V. Montalbano, Pisa Preprint IFUP Th-20/86 (unpublished);}

{C. Pittori and M. Testa, Zeitschrift fuer Physik C (Particles and Fields)
47 (1990) 487;}

{[6] L.D. Faddeev and S.L. Shatshvili, Phys. Lett. B167 (1986) 225;
L.D. Faddeev, Nuffield Workshop Proc. (1985);}

{[7] K. Harada and I. Tsutsui, Phys. Lett. B183 (1987) 311;}

{[8] O. Babelon, F.A. Shaposnik and C.M. Vialet, Phys. Lett. B177 (1986)
385;}

{[9] K. Fujikawa, Phys. Rev. Lett. 42 (1979) 11195;}

{[10] C. Becchi, A. Rouet and R. Stora, Ann. of Phys. 98 (1976) 287;
C. Becchi, A. Rouet and R. Stora, {\it Comments on Gauge Fixing II};}

{[11] T. Kugo and I. Ojima, Suppl. Prog. Theor. Phys. 66 (1979) 1;}

{[11A] Jordi Paris, Phys. Lett. B300 (1993) 104;

{[12] N. Nakanishi, Prog. Theor. Phys. 35 (1966) 1111;
B. Lautrup, Kgl. Danske. Videnskab. Selskab., Mat. Fis. Medd. 35 (1967) 1;}

{[13] A. M. Polyakov, Mod. Phys. Lett. A2 (1987) 893; V. G. Knizhnik, A. M.
Polyakov and A. B. Zamolodchikov, Mod. Phys. Lett. A3 (1988) 819;}

{[14] J. Distler and H. Kawai, Nucl. Phys. B321 (1989) 509;
F. David, Mod Phys Lett. A3 (1988) 1651}

{[15] L. Baulieu, C. Becchi and R. Stora, Phys. Lett. B180 (1986) 55;
L. Baulieu and M. Bellon, Phys. Lett. B196 (1987) 142;
C. Becchi, Nuc. Phys. B304 (1988) 513; see also K. Fujikawa, Nucl. Phys. 
B291 (1987) 583; K. Fujikawa, T. Inagaki and H. Suzuki, Phys. Lett. B213 
(1988) 279;} 

{[16] L. Alvarez-Gaum\`e and E. Witten, Nucl. Phys. B234 (1984) 269;}

{[17] M. N. Sanielvici, G. W. Senenoff and Y. Shi-Wu, Phys. Rev. Lett. 60 (1988)
2571}

{[18] M. Martellini, M. Spreafico and K. Yoshida, Mod. Phys. Lett. A7 (1992) 
1281; M. Martellini, M. Spreafico and K. Yoshida, Proc. International Workshop
of String Theory (21/26-9-1992) Accademia dei Lincei, Roma;}

{[19] M. Martellini, M. Spreafico and K. Yoshida, Mod. Phys. Lett. A9 (1994)
2009; M. Martellini, M. Spreafico and K. Yoshida, Rome Preprint 5/1994;}

{[20] H. Kawai, Y. Kitazawa and M. Ninomiya, Nucl. Phys. B393 (1993) 280, 
Nucl. Phys. B404 (1993) 684, hep/th 9511217;}

{[21] H. Kawai and R. Nakayama, Phys. Lett. B306 (1993) 224;}

{[22] T. Burwick, Nucl. Phys. B418 (1994) 257;}

{[23] J. Cohn and V. Periwal, Phys. Lett. B270 (1991) 18;}

{[24] D. Cangemi, R. Jackiw and B. Zwiebach, "Physical States in Matter-Coupled
Dilaton Gravity" hep/th 9505161;}

{[25] R. Jackiw, "Another View on Massless Matter-Gravity Fields in Two
Dimensions" hep/th 9501016; D. R. Karakhanyan, R. P. Manvelyan
and R. L. Mkrtchyan, Phys. Lett 329B (1994) 185; M. Martellini, Ann. of Phys.
(N.Y.) 167 (1986) 437;}

{[26] see eg. R. Jackiw, "Quantal Modifications to the Wheeler De Witt 
Equation" hep/th 9506037;}

{[27] J. Navarro-Salas, M. Navarro and C. F. Talavera, "Weyl Invariance
and the Balck Hole Equation" hep/th 9505139; G. Amelino-Camelia and D. 
Seminara, "Black Hole Radiation (with and) without Weyl Anomaly" MIT
Preprint MIT-CTP-2443.}

\vfill
\endpage
\end